\documentclass[aps,prd,aps,groupedaddress,preprintnumbers,nofootinbib,twocolumn,superscriptaddress,tightenlines]{revtex4-2}

\usepackage{amsmath, amssymb}
\allowdisplaybreaks
\usepackage{mathtools}
\usepackage[pdftex]{graphicx}
\usepackage[colorlinks]{hyperref}
\usepackage{color}
\usepackage{setspace, enumitem}
\usepackage[normalem]{ulem}
\usepackage{enumitem}
\usepackage{physics}
\usepackage[dvipsnames]{xcolor}
\usepackage{simpler-wick}
\usepackage{float}
\usepackage{amsmath}
\usepackage{hyperref}
\usepackage[capitalise]{cleveref}
\usepackage{youngtab}
\usepackage{ytableau}
\usepackage{multirow}
\usepackage{soul}

\ytableausetup{centertableaux}

\newcommand{\Nc}{N_C}
\newcommand{\Nf}{N_F}

\newcommand{\Sp}{\text{Sp}}
\newcommand{\SU}{\text{SU}}
\newcommand{\SO}{\text{SO}}
\newcommand{\U}{\text{U}}

\definecolor{HMG}{RGB}{212,175,55} 
\definecolor{BSB}{RGB}{4, 55, 242} 
\definecolor{JWG}{RGB}{153, 50, 204} 
\definecolor{GSG}{RGB}{243, 112, 243} 
\definecolor{AEG}{RGB}{31, 156, 85} 

\begin{document}
\title{\boldmath Dynamics of $E_6$ Chiral Gauge Theories}

\author{Andrew Goh}
\email{agoh@berkeley.edu}
\affiliation{Theory Group, Lawrence Berkeley National Laboratory, Berkeley, CA 94720, USA}
\affiliation{Berkeley Center for Theoretical Physics, University of California, Berkeley, CA 94720, USA}

\author{Hitoshi Murayama}
\email{hitoshi@berkeley.edu} 
\thanks{Hamamatsu Professor}
\affiliation{Theory Group, Lawrence Berkeley National Laboratory, Berkeley, CA 94720, USA}
\affiliation{Berkeley Center for Theoretical Physics, University of California, Berkeley, CA 94720, USA}
\affiliation{Kavli Institute for the Physics and Mathematics of the Universe (WPI),
University of Tokyo, Kashiwa 277-8583, Japan}

\author{Gup Singh}
\email{gupsingh@berkeley.edu}
\affiliation{Theory Group, Lawrence Berkeley National Laboratory, Berkeley, CA 94720, USA}
\affiliation{Berkeley Center for Theoretical Physics, University of California, Berkeley, CA 94720, USA}

\author{Bethany Suter}
\email{bethany\_suter@berkeley.edu}
\affiliation{Theory Group, Lawrence Berkeley National Laboratory, Berkeley, CA 94720, USA}
\affiliation{Berkeley Center for Theoretical Physics, University of California, Berkeley, CA 94720, USA}

\author{Jason Wong}
\email{jtwong71@berkeley.edu}
\affiliation{Theory Group, Lawrence Berkeley National Laboratory, Berkeley, CA 94720, USA}
\affiliation{Berkeley Center for Theoretical Physics, University of California, Berkeley, CA 94720, USA}

\begin{abstract}
    We present exact non-perturbative solutions to chiral gauge theories based on the $E_6$ gauge group and several matter fermions in the fundamental ${\bf 27}$-dimensional representation. They are obtained when supersymmetric versions are perturbed by small supersymmetry breaking by anomaly mediation. The universality classes obtained are very different from what can be conjectured by the tumbling hypothesis. In particular, the case with three ${\bf 27}$s may have an unbroken $\SU(3)$ symmetry with massless composite fermions in ${\bf 10}$ of $\SU(3)$. For this case, we employed numerical techniques to obtain the exact ground state.
\end{abstract}

\maketitle

\section{Introduction}
\label{sec:intro}

Nature is chiral, namely, it distinguishes left from right. The best theory for fundamental physics today is the so-called Standard Model of particle physics, which is a chiral gauge theory. Yet surprisingly, even the definition of a chiral gauge theory is not completely clear \cite{Ball:1988xg}. For instance, a standard gauge-invariant regularization method, such as Pauli--Villars with massive fields with wrong statistics to cancel UV divergences, is not possible for chiral gauge theories because the gauge invariance forbids a mass term.\footnote{A potential exception to the rule is an infinite tower of Pauli--Villars fields \cite{Narayanan:1993zzh}.} The dimensional regularization does not allow for a definition of the Levi--Civita tensor and hence one cannot define $\gamma_5 = \frac{i}{4!} \epsilon_{\mu\nu\rho\sigma} \gamma^\mu \gamma^\nu \gamma^\rho \gamma^\sigma$ as is required for a chiral projection. Lattice gauge theories have an intrinsic fermion doubling problem despite progress \cite{Ginsparg:1981bj,Kaplan:1992bt,Shamir:1993zy,Narayanan:1994gw,Luscher:1998pqa,Golterman:2000hr,Golterman:2004qv,Grabowska:2015qpk,Grabowska:2016bis}. Higher-derivative regularization (see, {\it e.g.}\/, \cite{Luscher:2000hn}) is not practical in most contexts.

Non-perturbative dynamics of chiral gauge theories is even less clear given the lack of their precise definition and lattice simulations to date. The only known way for an ``educated guess'' is the tumbling hypothesis. It postulates that fermions form bilinear condensates in the Most Attractive Channel (MAC) similar to QCD \cite{Raby:1979my}. Yet by definition of the theory being chiral, such bilinear operators are not gauge invariant. Given Elitzur's theorem, gauge-invariant operators cannot have a non-vanishing expectation value \cite{Elitzur:1975im}. Nonetheless, the hypothesis assumes exactly such a condensate that then breaks the gauge group by its own dynamics. By repeating this process, eventually the theory becomes vector-like, and the chain ends with a QCD-like condensate. This hypothesis identifies an unbroken global symmetry as well as a spectrum of massless composite fermions that match the 't Hooft anomalies in the original UV theory. Even though the method is applicable to all chiral gauge theories, it lacks theoretical justifications or evidence from numerical simulations.

Recently, it was pointed out that chiral gauge theories augmented by supersymmetry (SUSY) and anomaly-mediated supersymmetry breaking (AMSB) allow for exact solutions as long the size of supersymmetry breaking is small compared to the dynamical scale of the gauge theory $m \ll \Lambda$ \cite{Csaki:2021xhi}. This paper marked the first time that chiral gauge theories could be solved exactly. It is fascinating that the solution turned out to be very different from what was expected from the tumbling hypothesis. The challenge of this method is that supersymmetric gauge theories need to be studied individually for different gauge groups and particle contents, and it is much harder to draw general lessons compared to tumbling. It is imperative to accumulate a large number of cases to understand the general consequence of chiral gauge theories.

AMSB has the remarkable property of ``UV insensitivity" which grants predictive power for strongly coupled theories in four dimensions \cite{Randall_1999,Giudice:1998xp}. Namely, the effects of broken supersymmetry can be precisely studied in the IR theory regardless of UV physics. Furthermore, the supersymmetric limit is continuously connected to non-supersymmetric gauge theories via the supersymmetry breaking parameter $m$ which decouples squarks and gauginos. These noteworthy features of AMSB allow us to simply draw from known non-perturbative results in the pure supersymmetric theory to compute exact vacuum solutions in non-supersymmetric strongly coupled theories.

This program aims for a general tool to study strongly coupled chiral or non-chiral gauge theories by first studying their supersymmetric counterparts with small AMSB. Recently, this technique of perturbed AMSB has been applied to study a variety of chiral gauge theories \cite{Csaki:2021xhi, Csaki:2021aqv, Kondo:2022lvu, Leedom:2025mcg}. The vacuum structure and mass spectrum for these cases have been solved exactly. The technique has also been applied to Seiberg's results in supersymmetric theories \cite{Seiberg:1994bz, Seiberg:1994pq} to study the dynamics of QCD-like theories \cite{Murayama_2021, Kondo:2021osz, Csaki:2022cyg}. If we expect that the exact solutions for small SUSY breaking $m \ll \Lambda$ continuously connect to the non-SUSY limit $m \gg \Lambda$ where the theory is strongly coupled, we can uniquely identify the unbroken symmetries of the ground states and associated massless spectra of particles in non-SUSY gauge theories. In many cases, the near-SUSY limit $m \ll \Lambda$ allows for a weakly-coupled description, while the non-SUSY limit $m\gg \Lambda$ is intrinsically strongly-coupled and does not have well-controlled approximation methods. 

The BCS--BEC crossover, established both experimentally and theoretically in AMO and condensed matter physics \cite{Parish_2014,Randeria:2013kda}, provides an encouraging support for this program. The BCS (Bardeen--Cooper--Schrieffer) state due to a weak attractive force in a Fermi gas continuously connects to the BEC (Bose--Einstein Condensate) of bosonic bound states of fermions formed by non-perturbative strong attractive force. Knowing that it is a crossover, we can learn a lot about the strong coupling limit (BEC) by studying the weak coupling limit (BEC).

However, it is not decisively known whether or not there is a phase transition when the SUSY breaking parameter is taken to be on the same order of the strong coupling scale $m \sim \Lambda$, but there are plausible arguments which suggest that there will not be a phase transition, implying a continuous connection to non-SUSY limits where $m \rightarrow \infty$. Ultimately, if we are to anticipate an advent of theoretical breakthroughs in lattice gauge theory in the near future, the non-perturbative results obtained by AMSB will have to be verified with lattice simulations.

In this paper, we study the dynamics of non-SUSY chiral gauge theories based on $E_6$ using the method of AMSB with the supersymmetry breaking parameter taken to be infinitesimal. The group structure of $E_6$ is quite non-trivial in comparison to the more commonly studied classical groups. However, it serves as an additional class of simple chiral gauge theories that is worthy of examination given that $E_6$ admits complex representations in which all of its representations are anomaly-free by construction. Therefore, we can naturally embed Weyl fermions in the 27-dimensional representation. 
\section{Supersymmetric $E_6$ with 27's}
\label{sec:E6}



In the UV, we consider chiral superfields $\psi_{i} \; (i = 1,\ldots,\Nf)$ that are in the 27-dimensional fundamental representation of $E_6$ and in the fundamental representation of the $\SU(\Nf)$ global symmetry. The \textbf{27} representation of $E_6$ can be expressed in terms of the maximal subgroup $\SU(3) \times \SU(3) \times \SU(3)$ in which we find the following decomposition,
\begin{equation}\label{eq:e6repsu3}
    \textbf{27} \rightarrow (\textbf{3}, \bar{\textbf{3}}, \textbf{1}) \oplus (\textbf{1},\textbf{3},\bar{\textbf{3}}) \oplus (\bar{\textbf{3}},\textbf{1},\textbf{3}).
\end{equation}
The \textbf{27} can then be expressed as three $3\times3$ matrices. 
In the IR, below the strong coupling scale $\Lambda$, the theory can be described in terms of the gauge invariant polynomials, which are identified as coordinates of the moduli space. These invariants can be constructed using the fully symmetric, cubic invariant primitive tensor $d_{\mu \nu \lambda}$, \footnote{The $E_6$ primitive tensor $d_{\mu\nu\lambda}$ is normalized so that every entry is either $0,\pm1$.}
\begin{align}\label{eq:Stensor}
    S_{(ijk)} &= d_{\mu \nu \lambda}\psi_{i}^{\mu}\psi_{j}^{\nu}\psi_{k}^{\lambda} , \\
    \label{eq:Ttensor}
    T_{ijk;lmn} &= D_{\mu \nu \lambda ; \xi \rho \sigma }\psi_{i}^{\mu}\psi_{j}^{\nu}\psi_{k}^{\lambda} \psi_{l}^{\xi} \psi_{m}^{\rho} \psi_{n}^{\sigma} ,
\end{align}
where the invariant $T$ only appears for $\Nf \geq 3$ and the rank 6 invariant tensor $D_{\mu \nu \lambda ; \xi \rho \sigma }$ is formed from anti-symmetric products of the cubic invariant
\begin{equation}
\begin{aligned}\label{eq:T_full_expression}
    D_{\mu \nu \lambda ; \xi \rho \sigma } &=\frac{1}{36}d_{\alpha \beta \gamma}(d^{\mu\xi \alpha}d^{\nu\rho \beta}d^{\lambda\sigma \gamma} - d^{\nu\xi \alpha}d^{\mu\rho \beta}d^{\lambda\sigma \gamma} \\
    &\hspace{3cm} - d^{\nu\rho \alpha}d^{\mu\xi \beta}d^{\lambda\sigma \gamma} + \ldots).
\end{aligned}
\end{equation}

Here, the Greek indices ($\mu, \nu, \lambda, \xi, \rho, \sigma  = 1, \ldots , 27$) are $E_6$ fundamental representation indices and the Latin indices ($i, j, k, l, m, n = 1, \ldots, \Nf$) are $\SU(\Nf)$ flavor indices. These gauge invariants polynomials can also be expressed in terms of Young tableau
\begin{align}\label{eq:ST_young}
    S_{ijk} = 
    \begin{ytableau}
        i & j & k
    \end{ytableau}\, ,\ T_{ijk;lmn} =
    \begin{ytableau}
        i & j \\ 
        k & l \\ 
        m & n \\
    \end{ytableau}
\end{align}
In this work, we discuss cases $\Nf = 1, 2, 3$. For $\Nf = 4$, the moduli space is modified by quantum corrections and it is difficult to make concrete conclusions with AMSB. We therefore leave this case for future work. As each fundamental field develops a vacuum expectation value (vev), the generic symmetry breaking pattern is then $E_6 \rightarrow F_4 \rightarrow \SO(8) \rightarrow \SU(3)\rightarrow 1$. In each case, we use gauge symmetry to rotate the fields to leave only the first $3\times3$ matrix nonzero, so that the D-flat configurations reside in the $(\textbf{3}, \bar{\textbf{3}}, \textbf{1})$ part of the representation.

The result is summarized in \cref{tab:symmetry}, together with other cases that have already been studied \cite{Csaki:2021xhi,Csaki:2021aqv,Leedom:2025mcg,Kondo:2022lvu}. In most cases, the conjectured symmetry breaking pattern based on the tumbling hypothesis is not reproduced by the analysis based on SUSY and AMSB. Details follow in the next sections.

\begin{table*}
\def\arraystretch{1.7}
\begin{center}
\begin{tabular}{| c | c || c | c || c | c |}
    \hline
    \multicolumn{2}{|c||}{UV theory} & \multicolumn{2}{|c||}{SUSY $+$ AMSB} 
    & \multicolumn{2}{|c|}{Tumbling} \\ \hline
    $\Nf$ & $G_{\rm global}$  & $H_{\rm global}$  & $m_f=0$ 
     & $H_{\rm global}$  & $m_f=0$ \\ \hline\hline
    \multicolumn{6}{|c|}{$\SU(\Nc)+A(\hspace{0.1em}{\tiny\Yvcentermath1 \yng(1,1)} \hspace{0.1em})+(\Nc-4+\Nf)\tilde{F}(\bar{\scalebox{0.5}{\yng(1)}})+\Nf F({\scalebox{0.5}{\yng(1)}} )$, $\Nc=$Odd \cite{Csaki:2021xhi,Leedom:2025mcg}} \\ \hline
    $0$ & $\SU(\Nc-4)\times \U(1)$ & $\Sp(\Nc-5)\times \U(1)$ 
    & $\scalebox{0.5}{\yng(1)}+1$  &  $\SU(\Nc-4)\times \U(1)$ 
    & $\scalebox{0.5}{\yng(2)}$\\ \hline
    $1$ & $\SU(\Nc-3)\times \U(1)^2$ & $\Sp(\Nc-3)\times \U(1)$ 
    & {$\scalebox{0.5}{\yng(1)}$} & $\SU(\Nc-4)\times \U(1)$ & {$\scalebox{0.5}{\yng(2)}$} \\ \hline
    $2$ & $\SU(\Nc-2)\times \SU(2) \times \U(1)^2$ & $\Sp(\Nc-3)\times \U(1)^2$ 
    & $\scalebox{0.5}{\yng(1)}$ & $\SU(\Nc-4)\times \SU(2) \times \U(1)$ & $(\scalebox{0.5}{\yng(2)},1)$ \\ \hline \hline
    \multicolumn{6}{|c|}{$\SU(\Nc)+A(\hspace{0.1em}{\tiny\Yvcentermath1 \yng(1,1)} \hspace{0.1em})+(\Nc-4+\Nf)\tilde{F}(\bar{\scalebox{0.5}{\yng(1)}})+\Nf F({\scalebox{0.5}{\yng(1)}} )$, $\Nc=$Even \cite{Csaki:2021xhi,Leedom:2025mcg}} \\ \hline
    $0$ & $\SU(\Nc-4)\times \U(1)$ & $\Sp(\Nc-4)$ 
    & none & $\SU(\Nc-4)\times \U(1)$ 
    & $\scalebox{0.5}{\yng(2)}$ \\ \hline
    $1$ & $\SU(\Nc-3)\times \U(1)^2$ & $\Sp(\Nc-4)\times \U(1)$ 
    & none & $\SU(\Nc-4)\times \U(1)$ & $\scalebox{0.5}{\yng(2)}$ \\ \hline
    $2$ & $\SU(\Nc-2)\times \SU(2) \times \U(1)^2$ & $\Sp(\Nc-4)\times \SU(2) \times \U(1)$ 
    & none & $\SU(\Nc-4)\times \SU(2) \times \U(1)$ & $(\scalebox{0.5}{\yng(2)},1)$ \\ \hline\hline
    \multicolumn{6}{|c|}{$\SU(\Nc)+S({\scalebox{0.5}{\yng(2)}})+(\Nc+4)\tilde{F}(\bar{\scalebox{0.5}{\yng(1)}})$ \cite{Csaki:2021aqv}}\\ \hline
    $0$ & $\SU(\Nc+4)\times \U(1)$ & $\SO(\Nc+4)$ 
    & none & $\SU(\Nc+4)\times \U(1)$ & {\tiny\Yvcentermath1 \yng(1,1)} \\ \hline\hline
    \multicolumn{6}{|c|}{$\SO(10)+\Nf \psi({\bf 16})$ \cite{Kondo:2022lvu}} \\ \hline
    $1$ & none & none & none & none & none \\ \hline
    $2$ & $\SU(2)$ & $\SU(2)$ & none & $\SO(2)$ & none \\ \hline
    $3$ & $\SU(3)$ & $\SO(3)$ & none & $\SO(3)$ & none \\ \hline \hline
    \multicolumn{6}{|c|}{$E_6+\Nf \psi({\bf 27})$ [This Work]} \\ \hline
    $1$ & none & none & none & none & none \\ \hline
    $2$ & $\SU(2)$ & none & none & $\SO(2)$ & none \\ \hline
    \multirow{3}{*}{$3$} & \multirow{3}{*}{$\SU(3)$} & $\SU(3)$ & ${\scalebox{0.5}{\yng(3)}}$ & \multirow{3}{*}{$\SO(3)$} & \multirow{3}{*}{none} \\ \cline{3-4}
     & & $\U(1)\times\U(1)$ & $S_{iii} (i=1,2,3)$ & & \\ \cline{3-4}
     & & none & none & & \\ \hline
\end{tabular}
\end{center}
\caption{Conjectured global symmetries and massless fermion content of the non-supersymmetric chiral gauge theories. Representations of massless fermions are shown in Young tableaux of unbroken non-abelian groups. The SUSY $+$ AMSB column shows the exact results valid for $m\ll \Lambda$ which extend continuously to $m \gg \Lambda$ if there is no phase transition at $m \sim \Lambda$. The Tumbling column shows the hypotheses predicted by a sequential self-breaking of the gauge group by fermion bilinear operators in the Most Attractive Channel (MAC). }
\label{tab:symmetry}
\end{table*}

For the $E_6$ theories with $\Nf$ {\bf 27}'s, the tumbling hypothesis suggests $\SU(\Nf)$ is broken to $\SO(\Nf)$ with no massless fermions. The MAC is $(\psi^{\{i,} \psi^{j\}})({\bf 27}^*)$ whose vev breaks $E_6$ to $\SO(10)$. Out of $\psi({\bf 27})$, the singlet pair condenses and ${\bf 10+16}$ remain. The next MAC is $(\psi^{\{i,}({\bf 10})\psi^{j\}}({\bf 10}))({\bf 1})$ which condenses without further breaking of the gauge group. Finally $(\psi^{\{i,}({\bf 16})\psi^{j\}}({\bf 16}))({\bf 10})$ condenses and breaks $\SO(10)$ to $\SO(9)$. The ${\bf 16}$ is then a real representation and the $\SO(9)$ theory is vector like. The theory is gapped and there are no massless fermions.

What we find in this work is very different from what tumbling suggests. For $\Nf=1$, there is no global symmetry to speak of. For $\Nf=2$,  tumbling suggests $\SU(2) \rightarrow \SO(2)$, while SUSY$+$AMSB suggests no global symmetry left.  The case $\Nf=3$ is interesting. We found three local minima, each with different symmetry breaking patterns. Any one of them could persist to the non-SUSY limit $m\rightarrow\infty$. In particular, the case $\Nf=3$ with unbroken $\SU(3)$ with massless composite fermions in the $S_{ijk} = 
    \begin{ytableau}
        i & j & k
    \end{ytableau}$ representation was totally unanticipated.

\section{Anomaly Mediation \label{sec:AMSB}}
With anomaly mediated supersymmetry breaking (AMSB), we infinitesimally break $\mathcal{N} = 1$ supersymmetry. All supersymmetry breaking is parametrized in terms of the Weyl compensator superfield $\Phi = 1 + \theta^2 m$ where $\theta$ is the usual Grassmann coordinate in superspace. All dimensionful fields in the supersymmetric Lagrangian are multiplied by the Weyl compensator, producing a new Lagrangian
\begin{equation}
    \mathcal{L}=\int d^4\theta\, \Phi^*\Phi\, K+\int d^2\theta\, \Phi^3W+\text{h.c.}
\end{equation}
where $K$ and $W$ are the K\"ahler potential and superpotential, respectively. 
In a conformal theory, this compensator field could be removed by a simple rescaling of all fields by $\psi \rightarrow \Phi^{-1}\psi$, leaving supersymmetry unbroken. However, once a mass scale is introduced in the theory, the Weyl compensator can not be fully removed by a rescaling. This introduces a new term proportional to the SUSY breaking scale $m$ into the tree level Lagrangian
\begin{equation}
    \mathcal{L}_{\rm tree} = m\left( \psi_i^\mu \frac{\partial W}{\partial \psi_i^\mu} - 3W\right) + \text{h.c.}
\end{equation}
in addition to the normal F-term and D-term scalar potentials. 

It is often useful to require that the theory live along the D-flat directions, or equivalently at points in the moduli space where the D-term potential vanishes. This occurs when the following condition is satisfied
\begin{equation}
    \begin{aligned}
        D^a &= \sum_i \phi_i^\dag T^a \phi_i = 0.
    \end{aligned}
\end{equation}

With the addition of the new AMSB contribution, the potential now has a non-supersymmetric vacuum and the squarks and gauginos gain non-zero vevs.

\section{$\Nf=1$ Case}
\label{sec:Nf1}
\begin{figure}[h]
    \centering
    \includegraphics[scale=.75]{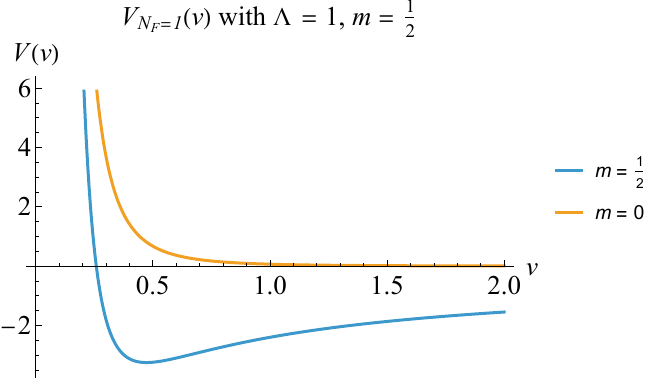}
    \caption{A plot of the $\Nf=1$ scalar potential as a function of $v$ with $\Lambda=1,$ and $m=1/2$. The plot shows the potential with AMSB $m=1/2$, and without AMSB $m=0$.}
    \label{fig:nf1_potential}
\end{figure}

For $E_6$ with one flavor, there is no continuous global symmetry. While there is a non-anomalous $\mathbb{Z}_6$ symmetry which acts on $\psi \rightarrow e^{2\pi i/6}\psi$. As long as $S \neq 0$, it is broken to $\mathbb{Z}_3$, which is identical to the center of the $E_6$ gauge group. Thus, there is no non-trivial global symmetry left. The $F_4$ gaugino condensate generates a nonperturbative superpotential for the theory. In terms of the IR field S, the superpotential takes the form
\begin{equation}
    W_{\Nf=1}=\left(\frac{\Lambda^{33}}{S^2}\right)^{1/9}.
\end{equation}
\indent A single D-flat direction breaks $E_6$ to $F_4$. Without loss of generality, we can restrict to the first $3 \times 3$ submatrix of the \textbf{27} of $E_6$. We use the $\SU(3) \times \SU(3)$ gauge symmetry to rotate the D-flat direction into the form
\begin{equation}
    \langle\psi\rangle=\begin{pmatrix}
        v&&\\&v&\\&&v
    \end{pmatrix}.
\end{equation}
In terms of $v$, the superpotential simplifies to
\begin{equation}
    \langle W_{\Nf=1}\rangle = \left(\frac{\Lambda^{33}}{6v^6}\right)^{1/9}.
\end{equation}
The total potential including the terms from AMSB (see \cref{fig:nf1_potential}) is
\begin{equation}
    \langle V_{\Nf=1}\rangle=\frac{2\times6^{5/9}\Lambda^{22/3}}{81v^{10/3}}-\frac{99\times6^{7/9}m\Lambda^{11/3}}{81v^{2/3}},
\end{equation}
which can be easily minimized to find the vev
\begin{equation}
    \langle v\rangle=\frac{2^{7/24}}{3^{5/6}}\left(\frac{5\Lambda^{11/3}}{11m}\right)^{3/8}.
\end{equation}
Thus, the minimum value of the scalar potential is
\begin{align}
    \langle V_{\Nf=1}\rangle_{\rm min}&=-\frac{33\times 3^{1/3}}{5\times2^{5/12}}\left(\frac{11m^5\Lambda^{11}}{4}\right)^{1/4}.
\end{align}
There is no global symmetry to check for the 't Hooft anomaly, so the mass spectrum only consists of the Higgs superfield which breaks supersymmetry. The corresponding fermion and scalar Higgs masses are
\begin{align}
    M_f^2&=\frac{121}{9}m^2,\ M_{b,\rm light}^2=\frac{242}{45}m^2,\ M_{b, \rm heavy}^2=\frac{968}{45}m^2,
\end{align}
which satisfies the supertrace rule.

\vspace{-5mm}
\section{$\Nf=2$ Case}
\label{sec:Nf2}

In this section, we discuss the features of $E_6$ with two fields in the fundamental representation. In addition to the gauge group $E_6$, there is also a global flavor symmetry $\SU(2)_F$. We expect the gauge symmetry to break  according to the following pattern
\begin{equation}
    E_6\rightarrow\SO(10)\times \U(1)_1\rightarrow\SO(8)\times \U(1)_1\times \U(1)_2.
\end{equation}
 The $\SO(8)$ gaugino condensate generates a non-perturbative superpotential of the form
\begin{equation}\label{eq:nf2_superpotential}
    W_{\Nf=2}=\left(\frac{\Lambda^{30}}{S^4}\right)^{1/6}.
\end{equation}
\begin{center}
\begin{table}[]
\begin{tabular}{| c | c | c | c |} 
 \hline
   & $X^+$ & $X^-$ & $\ Y\ $ \\ 
 \hline
 $\U(1)_1$ & $2$ & $2$ & $-4$ \\ 
 \hline
 $\U(1)_2$ & $1$ & $-1$ & $0$ \\
 \hline
\end{tabular}
\caption{Charges of the triplet of scalars in the embedding of $\SO(8)$ into a fundamental representation of $E_6$ for the case of $\Nf=2$ flavors. The gauge symmetry $\U(1)_1\times \U(1)_2$ is part of the 6-dimensional maximal torus of $E_6$.}
\label{tab:charges}
\end{table}
\end{center}
\vspace{-5mm}
We construct the $S^4$ global invariant from the standard normalization of the $\SU(2)_F$ Clebsch-Gordan coefficients. Since the only gauge invariant is the four-component $S$ tensor, the minimum must break the global $\SU(2)_F$ at least partially. In addition, the global symmetry suffers from the Witten anomaly as this theory has $27$ massless fermions in the UV. Thus, we expect the ground state produced from AMSB to fully break the global symmetry.

To come up with a general D-flat parameterization, we consider the symmetry breaking pattern discussed above. In breaking $E_6\to\SO(10)\times \U(1)_1$, the branching rule for $\textbf{27}$ of $E_6$ is that $\textbf{27}\rightarrow(\textbf{16},\textbf{-1})\oplus(\textbf{10},\textbf{2})\oplus(\textbf{1},\textbf{-4})$, so the embedding of $\SO(10)$ into the fundamental representation of $E_6$ is accomplished with a spinor, vector, and scalar of $\SO(10)$. Branching rules down to $\SO(8)$ gives us $\textbf{8}_v,\textbf{8}_s,\textbf{8}_c$ and three scalars. We can set the non-trivial representations of $\SO(8)$ to vanish in determining the general D-flat configuration, since the embedding of $\SO(8)$ into $E_6$ yields six $\SO(8)$ scalars in the two fundamental fields of $E_6$. The D-flat conditions from the remaining $\U(1)_1\times \U(1)_2$ fix two of these scalars, leaving four fields to parameterize our four D-flat directions. Each triplet of scalars under $\SO(8)$ has the charges given in \cref{tab:charges}.

The D-flat conditions for two pairs of these $\SO(8)$ singlets are then
\begin{align}
    2\left|X^+\right|^2+2\left|X^-\right|^2-4\left|Y^2\right|^2&=0,\nonumber\\\left|X^+\right|^2-\left|X^-\right|^2&=0,
\end{align}
where $X^\pm,Y$ are doublets of the global $\SU(2)_F$. Assuming $X^\pm,Y$ are real, the general solution is
\begin{align}
    X^+=v\begin{pmatrix}
        \cos\theta\\\sin\theta
    \end{pmatrix},\ X^-=v\begin{pmatrix}
        \cos\phi\\\sin\phi
    \end{pmatrix},\ Y=v\begin{pmatrix}
        \cos\chi\\\sin\chi
    \end{pmatrix}.
\end{align}
These scalars fit into the fundamental $E_6$ fields to produce the full D-flat configuration,
\begin{align}
    \langle\psi_1\rangle&=v\begin{pmatrix}
        \cos\theta&&\\&\cos\phi&\\&&\cos\chi
    \end{pmatrix},\nonumber\\\langle\psi_2\rangle&=v\begin{pmatrix}
        \sin\theta&&\\&\sin\phi&\\&&\sin\chi
    \end{pmatrix}.
\end{align}
Using the global $\SU(2)_F$ flavor symmetry, we rotate the field $\chi$ to 0 without loss of generality. 

In terms of this D-flat parametrization, the superpotential becomes
\begin{align}
    \langle W_{\Nf=2}\rangle&=\left(\frac{315}{7\sqrt{3}-3\sqrt{7}}\right)^{1/6}\times\nonumber\\&\times\frac{\Lambda^5}{v^2\left(\sin(\theta-\phi)\sin(\theta)\sin(\phi)\right)^{1/3}}.
\end{align}
The scalar potential thus takes the form
\begin{align}
    &\langle V_{\Nf=2}\rangle=\frac{35^{1/6}}{3\times 3^{2/3}}\left[\sqrt{2}\left(15(5+\sqrt{21})\right)^{1/6}\times\right.\nonumber\\&\left.\quad\quad\times\frac{\Lambda^{10}\left(2+\csc^2(\theta-\phi)+\csc^2(\theta)+\csc^2(\phi)\right)}{v^6\left(\sin(\theta-\phi)\sin(\theta)\sin(\phi)\right)^{2/3}}\right.\nonumber\\&\left.\quad-\frac{90m\Lambda^5}{(7\sqrt{3}-3\sqrt{7})^{1/6}v^2}(\sin(\theta-\phi)\sin(\theta)\sin(\phi))^{-1/3}\right].
\end{align}
We find a unique minimum where $\langle\theta\rangle=\pi/3$ and $\langle\phi\rangle=2\pi/3$, and 
\begin{equation}
    \langle v \rangle=\frac{2^{7/16}\left(7(5+\sqrt{21})\right)^{1/48}}{3^{1/16}\times5^{5/24}}\left(\frac{\Lambda^5}{m}\right)^{1/4}.
\end{equation}
The $\langle S_{111}\rangle$ and $\langle S_{122}\rangle$ components are non-vanishing in this case, breaking the global $\SU(2)_F$ fully. We also find no massless fermions in the spectrum in \cref{tab:masses_nf2} since the global symmetry fully breaks so the 't Hooft anomaly and Witten anomaly are trivially satisfied. \begin{table}[ht]
    \centering
    \def\arraystretch{1.7}
    \begin{tabular}{|c|c|c|c|c|}
        \hline
         &  Higgs  & $m_{(NG)}$  & $m_f = 0$ & $m_s$ \\
        \hline\hline
        \multicolumn{5}{|c|}{$\SU(2)_F\to1$} \\
        \hline
        $M_{b, \, \rm heavy}^2$ & $\frac{100}{3}m^2$ & $\frac{50}{9}m^2$ & $\varnothing$ & $\varnothing$ \\
        \hline
        $M_f^2$ & $25m^2$ & $\frac{25}{9}m^2$ & $\varnothing$ & $\varnothing$ \\
        \hline
        $M_{b, \, \rm light}^2$ & $\frac{50}{3}m^2$ & 0 & $\varnothing$ & $\varnothing$ \\
        \hline
        $1$ & $1$ & $3$ & $\varnothing$ & $\varnothing$ \\
        \hline
    \end{tabular}
    \caption{Masses of the D-flat directions for $\Nf=2$ flavors. We have divided up the masses of the scalars to those heavier than the fermions and lighter than the fermions, denoted by $M_{b, \, \rm heavy}^2$ and $M_{b, \, \rm light}^2$. Higgs refers to the supermultiplet in the field direction of the vev. The global $\SU(2)_F$ symmetry is fully broken, so that there are three Nambu-Goldstone bosons in the mass spectrum. In addition, there are no massless fermions which we attribute to the 't Hooft anomaly and Witten anomaly being trivially satisfied. The supertrace also vanishes as expected.}
    \label{tab:masses_nf2}
\end{table}
\begin{figure}[]
    \centering
    \includegraphics[scale=.8]{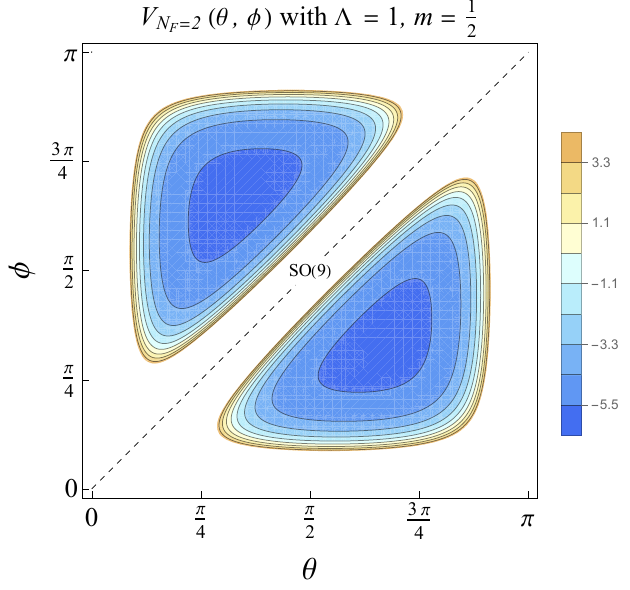}
    \caption{A plot of the $\Nf=2$ scalar potential over the angular coordinates $\theta,\phi$ with $\Lambda=1,$ and $m=1/2$. The dashed line indicates the region where the theory exhibits an enhanced $\SO(9)$ symmetry.}
    \label{fig:nf2_potential}
\end{figure}

\cref{fig:nf2_potential} shows the geometry of the scalar potential after introducing the AMSB term. In the angular $(\theta,\phi)$ space, the scalar potential is divided into triangular subregions bounded by enhanced $\SO(9)$ gauge symmetric parametrizations. The potential is periodic by $\pi$ in both $\theta$ and $\phi$, and exhibits a reflection symmetry $\theta\leftrightarrow\phi$. There is then a unique minimum at $\theta=\pi/2$ and $2\pi/3$, which corresponds geometrically to the centroid of the upper left triangle in \cref{fig:nf2_potential}.

In addition to the global $\SU(2)_F$ symmetry which is completely broken, there is a discrete ${\mathbb Z}_{12}$ symmetry where $\psi_{1,2} \rightarrow e^{2\pi i/12} \psi_{1,2}$. With $S \neq 0$, it is broken to ${\mathbb Z}_3$, which is identical to the center of the $E_6$ gauge group. Therefore there is no non-trivial global symmetry left.   
\vspace{-2mm}
\section{\boldmath $\Nf=3$ Case}
In this section, we discuss the case with a gauged $E_6$ and 3 fields in the fundamental representation, producing a global $\SU(3)_F$ flavor symmetry. In addition to the usual $S$ invariant used in the previous cases, there is also a new invariant $T$ as shown in \cref{eq:Ttensor}. The superpotential is now generated by the $\SU(3)$ gaugino condensate and in terms of $S$ and $T$, takes the form
\begin{equation}\label{eq:nf3_superpotential_text}
    W_{\Nf=3}=\left(\frac{\Lambda^{27}}{aT^3+bTS^4+cS^6}\right)^{1/3},
\end{equation}
where $a=1,b=6\sqrt{3},c=-28\sqrt{3}/5$ are coefficients determined by requiring the superpotential to blow up when the gauge symmetry doesn't fully break to $\SU(3)$ (see \cref{sec:nf3_superpotential_coefficients} for additional details). Using the Hilbert Series technique, we know that $S^4$ and $S^6$ are the only unique global invariants produced from the $S$ invariant. $T$ is a singlet of the global $\SU(3)_F$; a vacuum which only gives $T$ a vev would preserve the global symmetry, thereby requiring a non-trivial 't Hooft anomaly matching condition. 

There are eleven total D-flat directions which break $E_6$ down to $\SU(3)$. These match the eleven independent parameters found in the invariant polynomials, one in the singlet $T$ and ten in $S$. We describe a partial 10-dimensional parametrization of these D-flat directions in \cref{sec:10d}. We could not find a closed form 11-dimensional parameterization as the D-flat constraints yield a non-trivial set of algebraic equations. However we found that a 3-dimensional sub-manifold sufficed for finding possible ground states of the theory. Additional testing through numerical methods presented in \cref{sec:numerics} justifies this simplification. 

Since we are unable to find a full parameterization of the D-flat directions, we start with a single parameter that corresponds to the parameterizing the $T$ tensor. From here, we can add two more parameters that act as spherical parameters that tune $S^4$ and $S^6$. This parameterization is given below,
\begin{align}\label{eq:nf3_vev}
    \langle\psi_1\rangle&=v\begin{pmatrix}\cos\theta&&\\&\sin\theta\sin\phi&\\&&\sin\theta\cos\phi
    \end{pmatrix},\nonumber\\\langle\psi_2\rangle&=v\begin{pmatrix}
        &&-\sin\theta\sin\phi\\\sin\theta\cos\phi&&\\&\cos\theta&
    \end{pmatrix},\nonumber\\\langle\psi_3\rangle&=v\begin{pmatrix}
        &\sin\theta\cos\phi&\\&&\cos\theta\\\sin\theta\sin\phi&&
    \end{pmatrix}.
\end{align}
Unlike most theories in which SUSY is broken via AMSB, we find three vacua with different symmetry breaking patterns. Moreover, all of these cases satisfy 't Hooft anomaly matching conditions, so these minima are candidates in the non-SUSY limit. 

\begin{figure}[h]
    \centering
    \includegraphics[scale=.8]{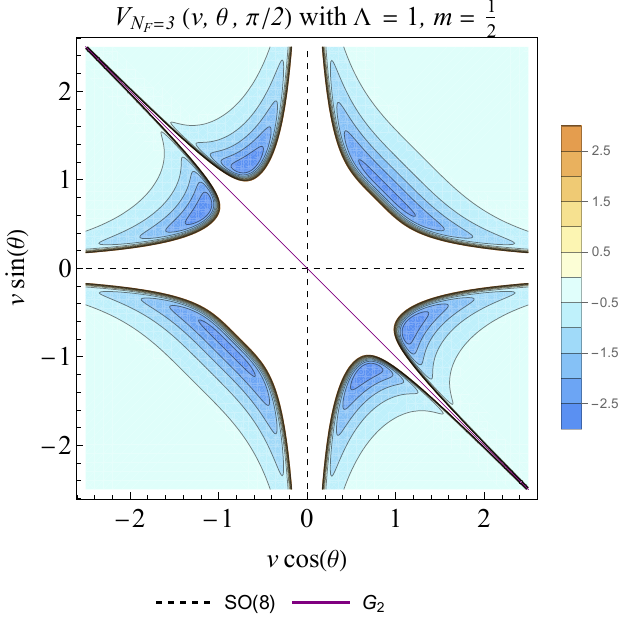}
    \caption{A plot of the $\Nf=3$ scalar potential over the coordinates $v,\theta$ with $\phi=\pi/2$, $\Lambda=1, \text{ and }  m=1/2$. The dashed and solid lines indicate regions of enhanced symmetry, splitting off the potential into separate regions that each contain a minimum.}
    \label{fig:nf3_potential_1}
\end{figure}

\begin{figure}[h]
    \centering
    \includegraphics[scale=.8]{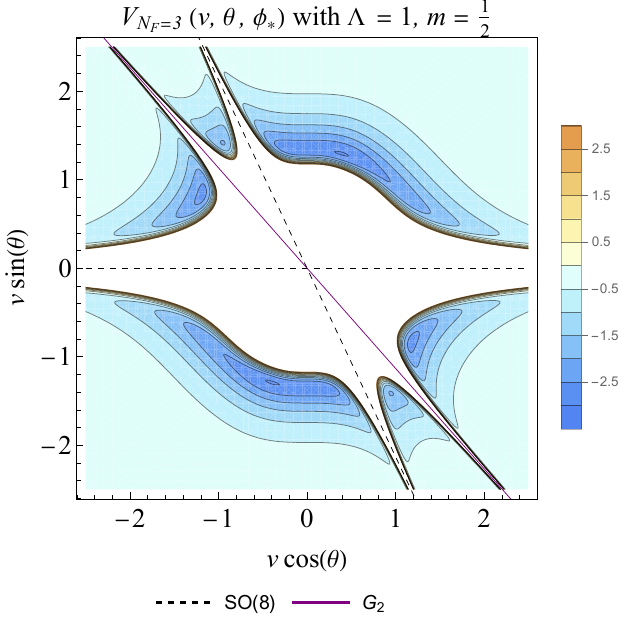}
    \caption{A plot of the $\Nf=3$ scalar potential over the coordinates $v,\theta$ with $\phi_*\approx0.57827$, $\Lambda=1,$ and $m=1/2$. The dashed and solid lines indicate regions of enhanced symmetry, splitting off the potential into separate regions that each contain a minimum. This particular cross section of the moduli space contains the deepest vacuum state with a value of around $V=-3.01811$.}
    \label{fig:nf3_potential_2}
\end{figure}

We ascribe the multiple distinct vacua to the structure of the superpotential in \cref{eq:nf3_superpotential_text}. The singular nature of the superpotential along enhanced symmetry regions corresponding to $\SO(8)$ and $G_2$ (see \cref{fig:nf3_potential_1} and \cref{fig:nf3_potential_2}) separates the scalar potential into three different regions. In each of these regions, AMSB can produce at most a single vacuum, and in our case of an $E_6$ gauge theory with $\Nf=3$ flavors of \bm{27}'s, we have found a maximal set of three valid vacua. 

\subsection{Unbroken Global $\SU$(3) Vacuum}
The vacuum that leaves the global symmetry fully unbroken corresponds to a vanishing $S$ invariant. Since the $T$ invariant is transforms trivially under the flavor symmetry when $\Nf=3$, a vev parameterized by $T$ can lead to a potential vacuum that leaves the global symmetry unaffected upon integrating out the broken gauge superfields. A non-tachyonic mass spectrum and matching 't Hooft anomaly further vindicates this vacuum as a valid minimum. 

We can achieve a vanishing $S$ invariant from \cref{eq:nf3_vev} by setting $\langle\theta\rangle=\pi/4$ and $\langle\phi\rangle=\pi/2$. Along this D-flat direction, the $T$ invariant is parameterized by the scale of the UV fields by $\langle T\rangle=v^6/2$. Since $\langle S^4\rangle=\langle S^6\rangle=0$ here, the superpotential takes a more convenient form
\begin{align}
    \langle W_{\Nf=3}\rangle|_T&=\frac{\Lambda^9}{a^{1/3}T}=\frac{2\Lambda^9}{a^{1/3}v^6}.
\end{align}
The scalar potential then has the explicit form
\begin{align}
    \langle V_{\Nf=3}\rangle|_T&=\frac{48\Lambda^{18}}{a^{2/3}v^{14}}-\frac{36m\Lambda^9}{a^{1/3}v^6},
\end{align}
which can be quickly minimized to obtain the vev
\begin{align}
    \langle v\rangle|_T&=\left(\frac{28\Lambda^9}{9a^{1/3}m}\right)^{1/8}.
\end{align}
In the limit $m\ll\Lambda$, the vev becomes parametrically large, so that our analysis starting with the UV fields is justified over working directly with the gauge invariants. The vacuum energy in this case is
\begin{align}
    \langle V_{\Nf=3}\rangle|_{T,\rm min}&=-\frac{108\sqrt{6}}{7^{7/4}a^{1/12}}\left(m^7\Lambda^9\right)^{1/4}\nonumber\\&=-8.78168\ldots\times\left(m^7\Lambda^9\right)^{1/4}.
\end{align}

At this vacuum, the global symmetry is left fully unbroken, so the mass spectra and 't Hooft anomaly should agree with this symmetry pattern. We find ten massless fermions in the spectrum, which correspond to the $S$ tensor due to the lack of $T$ in the superpotential along this direction. Indeed, the UV anomaly $\mathcal{A}_{\rm UV}=27$ matches the IR anomaly for the massless $\textbf{10}$ of $\SU(3)_F$ with $\mathcal{A}_{\rm IR}=(3+3)(3+6)/2=27$. 

\begin{table}[ht]
    \centering
    \def\arraystretch{1.7}
    \begin{tabular}{|c|c|c|c|c|}
        \hline
         &  Higgs  & $m_{(NG)}$  & $m_f = 0$ & $m_s$ \\
        \hline\hline
        \multicolumn{5}{|c|}{$\SU(3)_F\to\SU(3)_F, \quad V=-2.61081...$} \\
        \hline
        $M_{b, \, \rm heavy}^2$ & $\frac{648}{7}m^2$ & $\varnothing$ & 0 & $\varnothing$ \\
        \hline
        $M_f^2$ & $81m^2$ & $\varnothing$ & 0 & $\varnothing$ \\
        \hline
        $M_{b, \, \rm light}^2$ & $\frac{486}{7}m^2$ & $\varnothing$ & $0$ & $\varnothing$ \\
        \hline
        $\SU(3)_F$ & 1 & $\varnothing$ & ${\tiny\Yvcentermath1 \yng(3)}$ & $\varnothing$ \\
        \hline\hline 
        \multicolumn{5}{|c|}{$\SU(3)_F\to\U(1)_1\times\U(1)_2, \quad V=-2.61549...$} \\
        \hline
        $M_{b, \, \rm heavy}^2$ & $\frac{648}{7}m^2$ & $\frac{162}{49}m^2$ & $0$ & $\approx 15.53m^2$ \\
        \hline
        $M_f^2$ & $81m^2$ & $\frac{81}{49}m^2$ & $0$ & $\approx 11.22m^2$ \\
        \hline
        $M_{b, \, \rm light}^2$ & $\frac{486}{7}m^2$ & 0 & $0$ & $\approx 6.91m^2$ \\
        \hline
        $\U(1)_1\times\U(1)_2$ & $1$ & $6$ & $3$ & $1$ \\
        \hline\hline
        \multicolumn{5}{|c|}{$\SU(3)_F\to1, \quad V=-3.01811...$} \\
        \hline
        $M_{b, \, \rm heavy}^2$ & $\frac{648}{7}m^2$ & $\frac{162}{49}m^2$ & $\varnothing$ & $\frac{486}{49}m^2$ \\
        \hline
        $M_f^2$ & $81m^2$ & $\frac{81}{49}m^2$ & $\varnothing$ & $\frac{324}{49}m^2$ \\
        \hline
        $M_{b, \, \rm light}^2$ & $\frac{486}{7}m^2$ & 0 & $\varnothing$ & $\frac{162}{49}m^2$ \\
        \hline
        $1$ & $1$ & $8$ & $\varnothing$ & $2$ \\
        \hline
    \end{tabular}
    \caption{This table lists the masses of the D-flat directions as well as their representations under the global symmetries at tree level for each of the three vacua. For those with no non-abelian global symmetries remaining, the number instead denotes the multiplicity of the fields in each category. We have divided up the masses of the scalars to those heavier than the fermions and lighter than the fermions, denoted by $M_{b, \, \rm heavy}^2$ and $M_{b, \, \rm light}^2$. Higgs refers to the supermultiplet in the field direction of the vev. NG refers to Nambu--Goldstone supermultiplets for spontaneously broken global symmetries. Only the first two vacua have massless fermions; their scalar partners acquire positive mass squared from two-loop AMSB. The second vacua has one fermion and two scalars that were neither rational fractions nor proportional to common irrational numbers up to high numerical precision; these masses together fulfill the sum rule, a nontrivial test.}
    \label{tab:masses}
\end{table}

\subsection{Unbroken Global $\U$(1)$\times$$\U$(1) Vacuum}
The second vacuum is less trivial and breaks the global symmetry partially from $\SU(3)_F$ to $\U(1)_1\times\U(1)_2$. In this case, the global invariants $S^4$ and $S^6$ are algebraically dependent through the polynomial $P(S^4,S^6)=(S^4)^{1/4}-\alpha(S^6)^{1/6}$ for some constant $\alpha$. Moreover, the $T$ invariant is non-vanishing, so all terms in the denominator of the superpotential \cref{eq:nf3_superpotential_text} contribute towards the scalar potential. 

We can achieve a partially broken vacuum of $\U(1)_1\times\U(1)_2$ with a non-zero vev for only $\langle S_{123}\rangle$ of the $S$ invariant. With our parameterization in \cref{eq:nf3_vev}, we set $\langle\phi\rangle=\pi/2$ to produce this desired behavior. From our normalization convention described in \cref{sec:nf3_superpotential_coefficients}, the global invariants are parameterized by $v,\theta$ via
\begin{align}
    \left\langle S^4\right\rangle_{\phi=\pi/2}&=-\frac{v^{12}}{18\sqrt{3}}\left(\cos^3\theta-\sin^3\theta\right)^4,\nonumber\\\left\langle S^6\right\rangle_{\phi=\pi/2}&=-\frac{5v^{18}}{378\sqrt{3}}\left(\cos^3\theta-\sin^3\theta\right)^6,\nonumber\\\langle T\rangle_{\phi=\pi/2}&=\frac{v^6}{48}\left(10+6\cos4\theta+15\sin2\theta-5\sin6\theta\right).
\end{align}
Along these D-flat directions, the superpotential and scalar potential are
\begin{align}
    &\langle W_{\Nf=3}\rangle_{\phi=\pi/2}=\frac{\Lambda^92^{2/3}\csc^22\theta}{v^6\left(\cos^3\theta+\sin^3\theta\right)^{2/3}}\nonumber\\&\langle V_{\Nf=3}\rangle_{\phi=\pi/2}=\frac{\Lambda^{18}\csc^{16}\theta\sec^6\theta}{192\times2^{2/3}v^{14}\left(1+\cot^3\theta\right)^{10/3}}\times\nonumber\\&\quad\quad\times(55+12\cos4\theta-3\cos8\theta+24\sin2\theta-8\sin6\theta)\nonumber\\&\quad\quad-\frac{9m\Lambda^9}{2^{1/3}v^6}\frac{\csc^4\theta\sec^2\theta}{\left(1+\cot^3\theta\right)^{10/3}}.
\end{align}
With this form of the scalar potential, we can numerically minimize over the parameters $v,\theta$ to obtain a vacuum solution with
\begin{align}
    \langle v\rangle&=\left(5.06183\ldots\times\frac{\Lambda^9}{m}\right)^{1/8},\ \langle\theta\rangle=2.61549\ldots,
\end{align}
where $\langle\theta\rangle$ does not depend on the particular value of $m$ or $\Lambda$. The vacuum energy in this case has the form
\begin{align}
    \langle V_{\Nf=3}\rangle_{\rm min}&=-9.91804\ldots\times\left(m^7\Lambda^9\right)^{1/4}.
\end{align}
The corresponding mass spectra is shown in the second section of \cref{tab:masses} where we see that the supertrace rule is satisfied and that there are no tachyons in the spectrum. 

The less trivial check that this solution is a candidate vacuum is to show that the 't Hooft anomaly matches in the UV and IR. The unbroken global symmetry $\U(1)_1\times\U(1)_2$ correspond to the Cartan generators of $\SU(3)_F$. We have chosen our parameterization \cref{eq:nf3_vev} to already be diagonal under the $\U(1)_1\times\U(1)_2$ subgroup, so that our massless UV and IR fields transform as charges given in \cref{tab:nf3_anomaly}, from which we see that all of the 't Hooft anomalies match. 

\begin{table}[ht]
    \centering
    \begin{tabular}{|c | c | c | c | c | c | c|} 
         \hline
         & $\psi_1$ & $\psi_2$ & $\psi_3$ & $S_{111}$ & $S_{222}$ & $S_{333}$ \\ [0.5ex] 
         \hline
         $\U(1)_1$ & $0$ & $-1$ & $1$ & $0$ & $-1$ & $1$ \\ 
         \hline
         $\U(1)_2$ & $2$ & $-1$ & $-1$ & $6$ & $-3$ & $-3$ \\
         \hline
        \end{tabular}
    \caption{Charges of the unbroken $\U(1)_1\times\U(1)_2$ global symmetry for the massless UV and IR fermionic modes. }
    \label{tab:nf3_anomaly}
\end{table}
\vspace{-8mm}
\subsection{Fully Broken Global Symmetry Vacuum}

The final vacuum solution has vanishing $T$ and $S^4$ global invariants, fully breaking the global $\SU(3)_F$ symmetry. Thus, only the $S^6$ global invariant acquires a non-zero vev, and in this case we obtain the deepest scalar potential value at tree level for our choice of $m$ and $\Lambda$.

We numerically minimize (see \cref{sec:numerics}) the scalar potential over all three parameters $v,\theta,\phi$ in \cref{eq:nf3_vev} and obtain
\begin{align}
    \langle& v\rangle=\left(5.55590\ldots\times\frac{\Lambda^9}{m}\right)^{1/8},\nonumber\\\ \langle\theta\rangle&=1.27330\ldots,\ \langle\phi\rangle=2.05978\ldots
\end{align}
with a corresponding vacuum energy of
\begin{align}
    \langle V_{\Nf=3}\rangle_{\rm min}&=-10.15167\ldots\times\left(m^7\Lambda^9\right)^{1/4}.
\end{align}

The analytical expression for the global invariants, superpotential, and scalar potential are lengthy and can be found in \cref{app:globalinvariants}.

We observe no massless fermions and the 't Hooft anomaly is satisfied in this vacuum, since the $\SU(3)_F$ global symmetry of the theory is fully broken. The mass spectra obtained from our numerical analysis are shown in \cref{tab:masses}. Once again, we can see that the supertrace is satisfied, further verifying this as a candidate vacuum solution.

\section{Numerics}
\label{sec:numerics}
For $\Nf\leq2$ the relatively low dimensional moduli spaces allow for analytical methods. However in the case of $\Nf = 3$ with the introduction of the $T$ gauge invariant, we employ numerical techniques. We use the Python wrapper of \texttt{SymEngine}, a C++ backend symbolic manipulation library, and the Mathematica package \texttt{GroupMath} for the construction of our gauge invariants. We parse the \texttt{GroupMath} output using \texttt{ReGex} and create symbolic expressions in Python to differentiate and write our superpotential and thus our scalar potential.

To find an explicit expression for the $T$ invariant, we use the definition in \cref{eq:T_full_expression} in terms of the primitive invariant $d_{\mu\nu\lambda}$ of $E_6$, whose definition we use from \texttt{E6Tensors} \cite{Deppisch:2016xlp}. We can then export this tensor as a \texttt{CSV} file into Python and continue our analysis with \texttt{SymEngine}.

\indent We use the Broyden–Fletcher–Goldfarb–Shanno (BFGS) algorithm to minimize the scalar potential. To reduce computational complexity, we restrict the number of chiral superfields to the first nine components of each of our UV fields as described in \cref{sec:E6}. Minimizing our scalar potential gives us numerical values for these UV fields. From here, we substitute this numerical vev into the symbolic second derivative expression of the superpotential to obtain the fermion masses.

The complexity of our gauge invariants made obtaining scalar and pseudoscalar masses computationally challenging, so we outline a more efficient method. In this approach, we construct the superpotential and scalar potential by treating the global invariants as \texttt{SymEngine} functions of the UV fields. This enables us to treat the superpotential and scalar potential as functionals of the global invariants and differentiate implicitly twice with respect to the UV fields. Once these symbolically differentiated expressions are obtained, we evaluate the global invariants and their derivatives along a given vev. We can reduce the amount of computations further by invoking the commutativity of derivatives. Finally, we can substitute the vev into the second derivative of the scalar potential, where again we utilize the symmetric exchange of derivatives to decrease the computational steps. 

We found this approach to be significantly more efficient compared to writing the superpotential and scalar potential explicitly in terms of the UV fields. The structure of group invariants for linear representations are polynomials, so that derivatives of these global invariants only involve the power rule. Hence, the large number of $818100$ non-vanishing terms of the $T$ invariant does not pose an issue with runtime. The remaining derivatives only involve the global invariants, which takes minimal computation time.

With this technique, we obtain numerical mass spectra which appear in \cref{tab:masses}. We verify the analytical value of these masses to 512 binary bits of precision which correspond roughly to 154 decimals of precision. We perform the same numerical precision in all parts of our numerical calculations including the supertrace check. This computational approach to obtaining numerical solutions can be extended to other gauge theories as a useful tool alongside analytical methods, especially when the gauge theory and matter representations pose a high-dimensional problem that makes full analytical techniques difficult to implement.

We also include a short animation of the $\Nf=3$ potential that interpolates between \cref{fig:nf3_potential_1} and \cref{fig:nf3_potential_2} in an online repository \cite{full_repository}. We see that from these three D-flat directions in \cref{eq:nf3_vev} there are indeed only three distinct minima for this configuration, which are split by the enhanced symmetry regions corresponding to unbroken $\SO(8)$ and $G_2$ gauge symmetry.

\section{Conclusions}
\label{sec:conc}
In this paper, we obtained exact solutions to chiral $E_6$ gauge theories with $N_F=1,2,3$ matter fields in the {\bf 27} representation when supersymmetry is perturbed by a small supersymmetry breaking $m \ll \Lambda$ via anomaly mediation. The main results are summarized in Table~\ref{tab:symmetry}. Combined with all other results obtained using this method to date, we find quite different patterns of symmetry breaking and massless composite spectra from what could be conjectured based on the tumbling hypothesis in the non-supersymmetric limit. 

Even though there is a wealth of evidence that the exact solutions for $m \ll \Lambda$ smoothly connect to the non-supersymmetric limit $m \gg \Lambda$ for vector-like gauge theories \cite{kondo:2025nea}, there could still be phase transitions between the two limits in these chiral gauge theories. Nonetheless, we would like to emphasize that naively taking the limit $m \rightarrow \infty$ does give us a self-consistent conjecture on non-supersymmetric dynamics and it is worth exploring. 

Another important application is to stay within the validity range $m \ll \Lambda$ and use it to build models of physics beyond the standard model at high energies, where supersymmetry could indeed be present with a small supersymmetry breaking. In this case, we can employ the exact solutions we have directly, to address problems such as the composite axion for the Peccei--Quinn quality problem, dark matter with composite dynamics, composite inflaton field, non-perturbative dynamics for baryogenesis, etc. 

Finally, it would be useful to extend the analysis to more general types of supersymmetry breaking. We would like to give masses to both gauginos and scalars to approach the non-supersymmetric theory and the anomaly mediation provides them in a UV-insensitive fashion. On the other hand, D-terms [both conventional $\U(1)$ as well as $\U(1)_R$] are also UV-insensitive, while they do not generate gaugino masses and hence they are not sufficient for the purpose. Yet one can study the combination of them to see if there is an interesting phase diagram to be explored. In addition, some theories can incorporate a tree-level superpotential to explicitly break (a part of) the symmetries that could also provide a rich phase diagram to be studied. We hope to explore both of these directions in the near future to understand the big picture on dynamics of chiral gauge theories.

\begin{acknowledgments}
The work of B.A.S. is supported by the NSF GRFP Fellowship and BSF-2018140.
This material is based upon work supported by the National Science Foundation Graduate Research Fellowship Program under Grant No.\ DGE 2146752. Any opinions, findings, and conclusions or recommendations expressed in this material are those of the author(s) and do not necessarily reflect the views of the National Science Foundation. The work of H.M. is supported by the Director, Office of Science, Office of High Energy Physics of the U.S. Department of Energy under the Contract No. DE-AC02-05CH11231, by the NSF grant PHY-2210390, by the JSPS Grant-in-Aid for Scientific Research JP23K03382, MEXT Grant-in-Aid for Transformative Research Areas (A) JP20H05850, JP20A203, Hamamatsu Photonics, K.K, and Tokyo Dome Corportation. In addition, H.M. is supported by the World Premier International Research Center Initiative (WPI) MEXT, Japan.
\end{acknowledgments}
\clearpage
\onecolumngrid{}
\appendix

\section{Fixing Superpotential Coefficients for $\Nf = 3$}
\label{sec:nf3_superpotential_coefficients}

The superpotential generated by the $\SU(3)$ gaugino condensate is of the form
\begin{equation}\label{eq:nf3_superpotential}
    W_{\Nf=3}=\left(\frac{\Lambda^{27}}{aT^3+bTS^4+cS^6}\right)^{1/3},
\end{equation}
where $a,b,c$ are coefficients that we determine from the condition that the superpotential is poorly behaved when the gauge symmetry does not fully break to $\SU(3)$. 

We determine the $a,b,c$ relative coefficients using two different D-flat configurations that preserve a larger gauge symmetry than $\SU(3)$. One configuration that breaks $E_6$ to $G_2$ is the following,
\begin{equation}
    \langle\psi_1\rangle=\begin{pmatrix}
        1&&\\&0&\\&&0
    \end{pmatrix},\langle\psi_2\rangle=\begin{pmatrix}
        0&&\\&1&\\&&0
    \end{pmatrix},\langle\psi_3\rangle=\begin{pmatrix}
        0&&\\&0&\\&&1
    \end{pmatrix}.
\end{equation}
The $G_2$ configuration have non-vanishing $\langle T\rangle$ and $\langle S_{123}\rangle$ components so that we obtain a non-trivial relation in setting the denominator of the superpotential \cref{eq:nf3_superpotential} to vanish. 

A less trivial configuration that breaks $E_6$ to $\SO(8)$ is
\begin{equation}
    \langle\psi_1\rangle=\begin{pmatrix}
        u&x&x\\&&\\&&
    \end{pmatrix},\langle\psi_2\rangle=\begin{pmatrix}
        &&\\v&y&y\\&&
    \end{pmatrix},\langle\psi_3\rangle=\begin{pmatrix}
        &&\\&&\\w&z&z
    \end{pmatrix},
\end{equation}
and where the D-flat constraints require that $y^2+u^2=x^2+v^2$ and $z^2+u^2=x^2+w^2$.

Our choice of normalization convention for $T,S^4,$ and $S^6$ is based on the \texttt{GroupMath} package. We generate the $T$ tensor through its definition in \cref{eq:Ttensor} including a factor of $1/36$ to account for the $36$ terms of products of $S$ to match the symmetry pattern in \cref{eq:ST_young}. The $S^4$ and $S^6$ invariants are produced from the canonical definition of $S$ in \cref{eq:Stensor} and plugged into \texttt{GroupMath} to yield $\SU(3)$ invariant tensors.

With these normalizations and testing these enhanced symmetry configurations, we obtain relative coefficients
\begin{align}
    b=6\sqrt{3}a,\ c=-\frac{28\sqrt{3}}{5}a,
\end{align}
and from now on we set $a=1$ as a convention. This exact value can be determined through an exact gaugino condensate calculation, but we find that this is not necessary for the rest of this computation.

\section{\boldmath 10D Parametrization of Moduli Space}
\label{sec:10d}
The case of $\Nf=3$ flavors of $E_6$ has eleven D-flat directions. Here, we describe how to write an explicit parameterization of ten of these D-flat directions. Let us start by embedding nine vectors in $\mathbb{R}^3$ labeled $\mathbf{l}_i,\mathbf{m}_j,\mathbf{n}_k$ with $i=1,2,3$ into the $(\mathbf{3},\bar{\mathbf{3}},\mathbf{1})$ block of three $\mathbf{27}$s of $E_6$ as
\begin{align}
    \langle\psi_i\rangle&=\begin{pmatrix}
        \mathbf{l}_1&\mathbf{m}_3&\mathbf{n}_2\\\mathbf{n}_3&\mathbf{l}_2&\mathbf{m}_1\\\mathbf{m}_2&\mathbf{n}_1&\mathbf{l}_3
    \end{pmatrix}.
\end{align}
We simplify our analysis by setting $\mathbf{l}_3,\mathbf{m}_3,\mathbf{n}_3$ to vanish. The D-flat conditions then restrict the form of the remaining vectors for $i\neq j$,
\begin{align}
    &\mathbf{l}_i\cdot\mathbf{m}_j=\mathbf{m}_i\cdot\mathbf{n}_j=\mathbf{n}_i\cdot\mathbf{l}_j=0\nonumber\\\sum_{i=1}^2&\mathbf{l}_i\cdot\mathbf{l}_i=\sum_{i=1}^2\mathbf{m}_i\cdot\mathbf{m}_i=\sum_{i=1}^2\mathbf{n}_i\cdot\mathbf{n}_i.
\end{align}
These constraints can be solved for analytically using Mathematica to produce a 10-dimensional parameterization of the D-flat directions for $E_6$ with three $\mathbf{27}$s. To do this, we set the length of $|\mathbf{l}_i|=|\mathbf{m}_i|=|\mathbf{n}_i|$ and then impose orthogonality conditions. We satisfy the orthogonality constraints by letting $\mathbf{l}_i$ be general vectors, and set the complement $\mathbf{m}_j$ and $\mathbf{n}_j$ with $i\neq j$ to span the orthogonal plane to $\mathbf{l}_i$. We can then solve the remaining orthogonality condition $\mathbf{m}_i\cdot\mathbf{n}_j=0$ explicitly in this reduced parameter space. This yields a 7-dimensional parameterization of the D-flat directions, but we can include three additional parameters corresponding to the global $\SU(3)_F$ symmetry.

Adding an eleventh parameter to the D-flat parameterization requires one of $\mathbf{l}_3,\mathbf{m}_3,\mathbf{n}_3$ to be non-zero. This greatly increases the number of orthogonality conditions, leading to equations that are much more difficult to separate and solve. We do not pursue a full parameterization because of this complication, and numerical checks improve our confidence that an entire parameterization of D-flatness is unnecessary for our analysis.

\section{Global Invariants for $\Nf = 3$ Fully Broken Global Symmetry Case\label{app:globalinvariants}}
In the $E_6$ gauge theory with $\Nf=3$ flavors, we can use the vev in \cref{eq:nf3_vev} to find the resultant global invariants, as well as the superpotential and scalar potential, explicitly. These expressions are analytical, but lengthy, so we have provided their form in this appendix. For brevity, we write $s_\theta=\sin\theta$ and $c_\theta=\cos\theta$. With these simplifications, the analytical expression for the global invariants are
\begin{align}
    \langle T \rangle &= \frac{2}{3} v^6 \Big(5 s^3_{2 \theta} s^3_\phi + s^6_{\theta} \left(5 s^3_{2 \phi} + 4 \left(s^6_\phi + c^6_\phi \right)\right) - 40 s^3_\theta c^3_\theta c^3_\phi + 4 c^6_\theta \Big) \nonumber\\
    \langle S^4 \rangle &=  \frac{v^{12}}{36 \sqrt{3}} \Big(512 s^9_\theta c^3_\theta \left(s^9_\phi - c^9_\phi + 57 s^3_\phi c^6_\phi - 57 s^6_\phi c^3_\phi \right) + 57 s^6_{2 \theta} s^3_{2 \phi} - \frac{1}{2} s^{12}_\theta \left(4 s^3_\phi - 3 c_\phi - c_{3 \phi }\right)^4 - 96 s^6_\theta c^6_\theta (3 c_{4 \phi} + 5)  \nonumber\\
    &\hspace{2 cm} - 512 s^3_\theta c^9_\theta \left(c^3_\phi -s^3_\phi \right) - 128 c^{12}_\theta \Big)\nonumber\\
    \langle S^6 \rangle &= \frac{1280 v^{18}}{189 \sqrt{3}} \bigg(\frac{15}{16} s^6_\theta c^{12}_\theta (17 s_{6 \phi} - 51 s_{2 \phi} - 6 c_{4 \phi} - 10) + 6 s^3_\theta c^{15}_\theta \left(s^3_\phi -c^3_\phi \right) + 20 s^9_\theta c^9_\theta \left(s^9_\phi - c^9_\phi - 78 s^3_\phi  c^6_\phi + 78 s^6_\phi c^3_\phi \right)\nonumber\\
    &\hspace{2 cm} +\frac{s^{18}_\theta}{4096} \left(4 s^3_\phi - 3 c_\phi - c_{3 \phi }\right)^6 - \frac{3}{256} s^{15}_\theta c^3_\theta \left(-4 s^3_\phi + 3 c_\phi + c_{3 \phi} \right)^3 (66 s_{2 \phi} - 22 s_{6 \phi} + 3 c_{4 \phi} + 5)\nonumber\\
    &\hspace{2 cm} - \frac{3}{512} s^{12}_\theta c^6_\theta (9360 s_{2 \phi} - 520 s_{6 \phi} - 1560 s_{10 \phi} - 6330 + 12465 c_{4 \phi} - 4326 c_{8 \phi} + 751 c_{12 \phi}) - c^{18}_\theta \bigg).
\end{align}
The superpotential takes a rather simple form,
\begin{align}
    \langle W\rangle&=\frac{2^{5/3} \Lambda ^9}{v^6s_\theta^2\left[\left(c_{\phi }^3+s_{\phi }^3\right)^2 \left(c_{\theta }^3-c_{\phi }^3 s_{\theta }^3\right)^2 \left(c_{\theta }^3+s_{\theta }^3 s_{\phi }^3\right)^2\right]^{1/3}}.
\end{align}
In terms of the superpotential, the scalar potential becomes:

\begin{align}
    \langle V\rangle &= -\frac{v^{16} \langle W\rangle^4 s_{\theta }^4}{48 \Lambda ^{27}} \bigg[c_{\theta }^{12} \left(27 m v^2 c_{\phi }^6
   s_{\theta }^2+54 m v^2 c_{\phi }^3 s_{\theta }^2 s_{\phi }^3-4 \langle W\rangle c_{\phi }^4+27
   m v^2 s_{\theta }^2 s_{\phi }^6-4 \langle W\rangle s_{\phi }^4\right)\nonumber\\
    &\hspace{3 cm} +2 c_{\theta }^9 s_{\theta
   }^3 \left(-27 m v^2 c_{\phi }^9 s_{\theta }^2-27 m v^2 c_{\phi }^6 s_{\theta }^2
   s_{\phi }^3+27 m v^2 c_{\phi }^3 s_{\theta }^2 s_{\phi }^6+8 \langle W\rangle c_{\phi }^7+27 m
   v^2 s_{\theta }^2 s_{\phi }^9-8 \langle W\rangle s_{\phi }^7\right)\nonumber\\
    &\hspace{3 cm}+c_{\theta }^6 s_{\theta }^6
   \bigg(8 \langle W\rangle c_{\phi }^4 s_{\phi }^6-16 \langle W\rangle s_{\phi }^{10}+2 c_{\phi }^6 \left(4 \langle W\rangle s_{\phi }^4-81 m v^2 s_{\theta }^2 s_{\phi
   }^6\right)+2 c_{\phi }^3 \left(8 \langle W\rangle s_{\phi }^7-27 m v^2 s_{\theta }^2 s_{\phi
   }^9\right)\nonumber\\
    &\hspace{5.5 cm}+16 \langle W\rangle c_{\phi }^7 s_{\phi }^3+27 m v^2 c_{\phi }^{12} s_{\theta }^2-16 \langle W\rangle c_{\phi
   }^{10}+27 m v^2 s_{\theta }^2 s_{\phi }^{12}-54 m v^2 c_{\phi }^9 s_{\theta }^2
   s_{\phi }^3\bigg)\nonumber\\
    &\hspace{3 cm}-2 c_{\theta }^3 c_{\phi }^3 s_{\theta }^9 s_{\phi }^3 \bigg(-27 m v^2 c_{\phi }^9
   s_{\theta }^2-27 m v^2 c_{\phi }^6 s_{\theta }^2 s_{\phi }^3+c_{\phi }^3
   \left(27 m v^2 s_{\theta }^2 s_{\phi }^6-8 \langle W\rangle s_{\phi }^4\right)+8 \langle W\rangle c_{\phi }^4
   s_{\phi }^3\nonumber\\
    &\hspace{5.5 cm}+16 \langle W\rangle c_{\phi }^7 +27 m v^2 s_{\theta }^2 s_{\phi }^9-16 \langle W\rangle s_{\phi
   }^7\bigg)\nonumber\\
    &\hspace{3 cm}+c_{\phi }^4 s_{\theta }^{12} s_{\phi }^4 \left(c_{\phi }^8 \bigg(27 m
   v^2 s_{\theta }^2 s_{\phi }^2-4 \langle W\rangle\right)+c_{\phi }^5 \left(54 m v^2 s_{\theta
   }^2 s_{\phi }^5-16 \langle W\rangle s_{\phi }^3\right)-16 \langle W\rangle c_{\phi }^6 s_{\phi }^2\nonumber\\
    &\hspace{5.5 cm}+c_{\phi }^2 \left(27 m v^2 s_{\theta }^2
   s_{\phi }^8-16 \langle W\rangle s_{\phi }^6\right)-16 \langle W\rangle c_{\phi
   }^3 s_{\phi }^5-4 \langle W\rangle s_{\phi }^8\bigg)\nonumber\\
    &\hspace{3 cm}-16 \langle W\rangle c_{\theta }^{10} s_{\theta }^2
   \left(c_{\phi }^3+s_{\phi }^3\right){}^2+16 \langle W\rangle c_{\theta }^7 s_{\theta }^5
   \left(c_{\phi }^3-s_{\phi }^3\right) \left(c_{\phi }^3+s_{\phi }^3\right){}^2-4
   \langle W\rangle c_{\theta }^4 s_{\theta }^8 \left(c_{\phi }^6-s_{\phi
   }^6\right){}^2\bigg].
\end{align}

\twocolumngrid{}
\bibliographystyle{utphys}
\bibliography{AMSB_Refs}

\end{document}